\documentclass{aa}
\usepackage{graphicx}
\usepackage{verbatim}
\usepackage{amssymb}
\usepackage{natbib}
\bibpunct{(}{)}{;}{a}{}{,}
\newcommand{\pas}{.\hskip-2pt$^{\prime\prime}$}
\titlerunning{Water masers in AFGL5142 with EVN}
\begin{document}
   \title{EVN observations of H$_2$O masers towards the high-mass young stellar object in AFGL 5142}

  \author{C. Goddi \inst{1}\fnmsep\inst{2} 
\and L. Moscadelli \inst{1} \and  W. Alef \inst{3} \and J. Brand \inst{4}}

   \offprints{C. Goddi,\\\email{cgoddi@ca.astro.it}}

   \institute{INAF, Osservatorio Astronomico di Cagliari, Loc. Poggio dei Pini,
 Str. 54, 09012 Capoterra (CA), Italy
   \and
      Dipartimento di Fisica, Universit{\`a} degli Studi di Cagliari,
       S.P. Monserrato-Sestu Km 0.7, I-09042 Cagliari, Italy
         \and
             Max-Planck-Institut f{\"u}r Radioastronomie, Auf dem 
H{\"u}gel 69,
             D-53121 Bonn, Germany
          \and 
            Istituto di Radioastronomia CNR, Via Gobetti 101, 40129 Bologna, Italy 
             }

   \date{Received ``date'' / Accepted ``date''}

   \abstract{

We have conducted multi-epoch EVN observations of the 22.2~GHz water masers towards the high-mass young stellar object in AFGL 5142. 
With four observing epochs, spanning a time of $\sim$1~year,
12 distinct maser features have been detected, 7 out of these 
detected in more than one epoch. 
The positions and the velocities of the VLBI features agree well with those
of the emission centers previously identified by means of VLA observations.
For a few features, persistent over three 
or four epochs, accurate values of the proper motions are derived. The 
observed proper motions have an amplitude of 15--20~km~s$^{-1}$, significantly
larger than the range of variation of the line-of-sight velocities 
($\pm$4~km~s$^{-1}$ around the systemic velocity).
On the basis of their spatial 
distribution, the observed maser features can be divided into two groups. 
A model fit to the positions and velocities of the maser features of Group I, 
detected in the same region (within $\sim$500~mas) where the 
massive YSO should be located, demonstrates that these might arise on the surface of a nearly edge-on Keplerian disk, rotating around a massive young stellar object.
The maser features of Group II, found at large distances from the YSO
($\geq$ 1$''$), have positions and line-of-sight velocities in agreement
with the blue-shifted lobe of a large scale molecular outflow (traced by the 
HCO$^{+}$ and SiO emission), and might 
result from the interaction between the gas flowing away from the young stellar object and the ambient gas of the progenitor molecular core.

   }

  \maketitle

\section{Introduction}
The star formation
process is  better understood  for low-mass stars ($\sim$1~M$_\odot$) than for high-mass stars ($\geq$ 10~M$_\odot$). The
massive stars are less numerous, on average more distant from the Sun, and  enter the ZAMS phase still enshrouded in their progenitor dust and gas envelope, making optical and near-infrared observations impracticable. This explains why, to date, only a handful of high-mass protostellar candidates 
have been identified \citep{Ces97, She98, Hun98, Mol98, Fon04}. To study the formation process of massive stars, high resolution observations at radio, millimetre and far-infrared
wavelenghts are needed. The highest resolutions ($\leq $ 1 mas) are obtainable through the  Very Long Baseline Interferometry (VLBI) technique at
radio wavelenghts, which can be used to observe the maser transitions of several molecular species, such as OH, H$_{2}$O, CH$_{3}$OH, observed in the proximity of the
high-mass proto-star.
Multi-epoch VLBI observations can provide accurate (relative) positions, line-of-sight and 
transversal velocities of the maser spots (the individual, mas-scale compact, centers of 
emission), so that the 3-dimensional velocity distribution of the gas traced by the maser 
transition can be derived. Moreover, the VLBI measurements of the emission properties of the single maser spots can be usefully compared with the maser excitation models to constrain the relevant physical 
and geometrical parameters. 

 The first multi-epoch VLBI experiments were performed about 20~years
 ago towards a few of the 
strongest 22.2~GHz H$_2$O interstellar masers in the Galaxy, 
i.e. Orion-KL \citep{Gen81a}, W51 \citep{Gen81b}, Sagittarius B2  \citep{Rei88}, 
W49 \citep{Gwi92}, and allowed both the determination of the characteristic pattern of the
kinematics of these regions and, by comparing the line-of-sight velocities with the proper motions
of the spots, the derivation of accurate source distances. Since then, VLBI 
observations of 22.2~GHz water masers have been mostly focused on those sources which,
selected on the basis of interferometrical observations of (thermal) molecular  tracers, 
are considered to be among the best high-mass protostellar candidates.
The Very Long Baseline Array (VLBA) observations carried out by \citet{Mos00} 
towards the water masers in IRAS 20126+4104, one of the best studied 
examples of a high-mass Young Stellar Object (YSO) associated with a Keplerian disk and a jet/outflow system \citep{Ces97}, suggest that the H$_2$O masers could 
arise on a
bipolar conical surface, excited by the interaction of an ionized jet with the surrounding molecular gas. VLBA observations of the 22.2~GHz masers performed towards intermediate-mass
(NGC 2071, \citealt{Set02}) and high-mass (IC~1396N, \citealt{Sly99};
W3 IRS 5, \citealt{Ima00}) YSOs indicate that this maser emission can originate both
in the protostellar disks and at the base of the molecular outflows. 

This paper presents multi-epoch VLBI observations of the 22.2~GHz H$_2$O masers 
towards the high-mass star forming region AFGL~5142 (IRAS05274+3345).  
At a distance of 1.8 kpc \citep{Sne88}, its far-infrared luminosity is estimated
to be \ $ 3.8 \times 10^3$ L$_\odot$ \citep{Car90}. 
Using the Very Large Array (VLA) at 8.4~GHz \citet{Tof95} have revealed a 
faint (1 mJy), almost 
unresolved, continuum source (best interpreted as free-free emission from an 
ionized wind), which subsequently has been found to be coincident in position 
with the center of a CO bipolar outflow and with the origin of a jet observed 
in the H$_2$ near-infrared emission \citep{Hun95}. Owens Valley Radio Observatory 
(OVRO) data by \citet{Hun99} show: \ 1) a well-collimated SiO-jet and an HCO$^+$-outflow, both aligned with the axis of the CO-outflow, and emanating from the 8.4 GHz 
continuum source; \ 2) a compact 88 GHz continuum source coincident in position
(within the observational errors) with the 8.4 GHz continuum. The radio
flux and the bolometric luminosity (estimated using the IRAS
fluxes) can be explained if the exciting source is a massive object, with spectral type
B2 or earlier.
The VLA NH$_3 $-observations performed by \citet{Zha02} show a 1$''$ diameter (1800 AU) compact structure, whose morphology and  kinematics are compatible with
 a rotating disk surrounding a high-mass young star. 

The 22.2~GHz water masers in AFGL~5142 have been observed with the VLA at three epochs 
(1991, \citealt{Tor92}; 1992, \citealt{Hun95}; 1998, \citealt{Hun99}), and found to be distributed within
a few arcseconds from the 8.4 GHz continuum source. However, the VLA angular
resolution ($\sim$0\pas1) is inadequate to determine the detailed spatial 
distribution of the maser spots, to measure their proper motions, and to 
investigate the kinematics traced by the H$_{2}$O masers in this source. 
At the distance of AFGL 5142, the 
linear resolution attainable using the VLBI technique at 22.2~GHz  
is $\sim$2~AU, sufficient in principle 
 to resolve and determine the velocity profile of a protostellar disk 
(of size of hundreds of AU) and/or to study the mechanism of collimation
and acceleration of a jet.

Section~2 of this paper describes our multi-epoch VLBI observations and gives
technical details of the data analysis, while Section~3 presents the observational results. 
Section~4 describes plausible kinematical models for interpreting the measured positions and 
velocities of the maser features.
The conclusions are drawn in Section~5.


\begin{figure}
\includegraphics[angle= -90, width=\hsize, trim=0.5cm 0cm 0cm 5cm, clip]{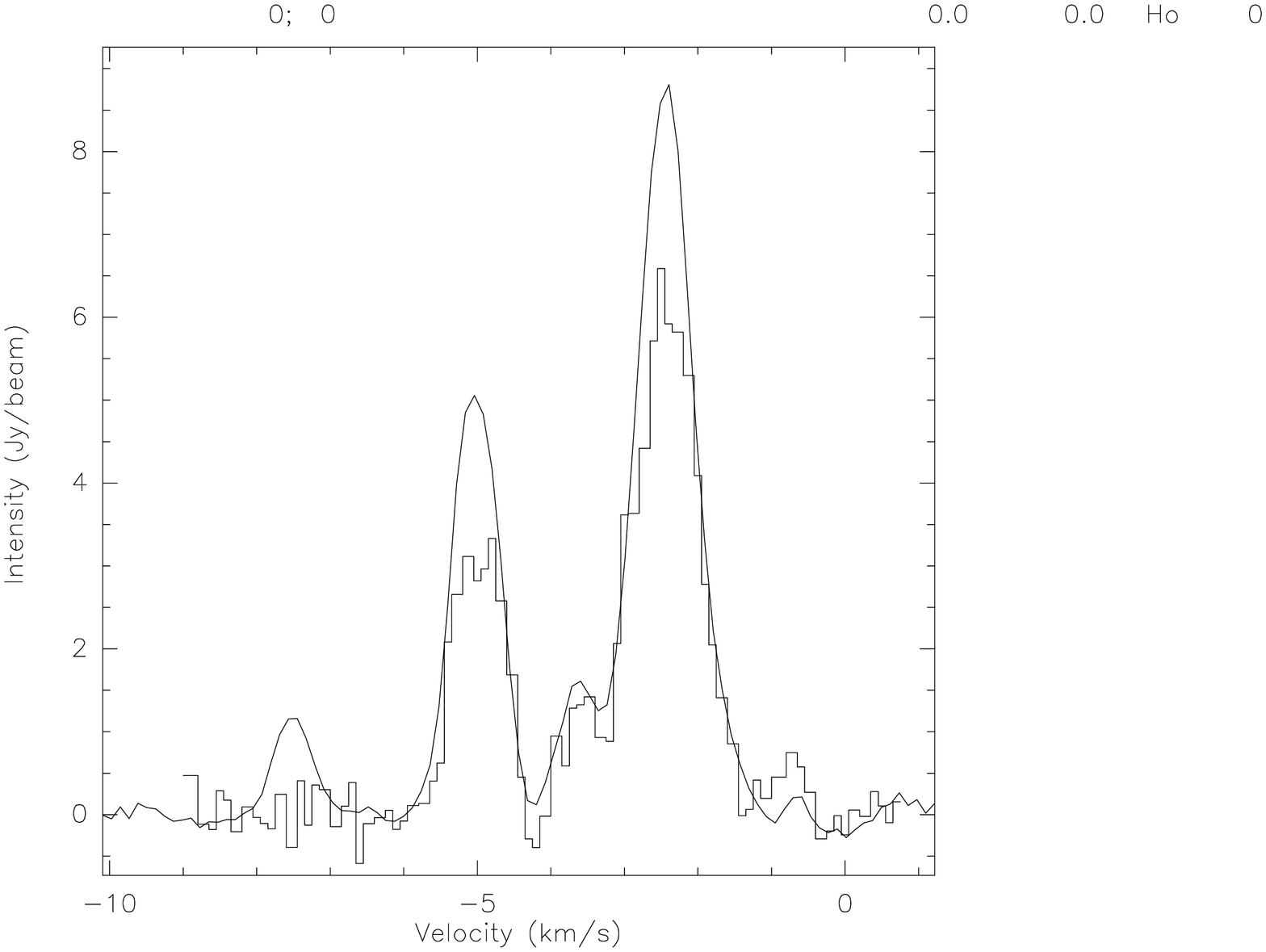}
\caption{The Effelsberg total-power spectrum observed in Oct 1996 (connected plot) is compared with the integrated flux densities (histogram plot) of the VLBI channel maps. The velocity resolution of both spectra is equal to the channel width of 0.12~km~s$^{-1}$.}

\label{spec}
\end{figure}
\section{Observations and data reduction}

\begin{table*}
\centering
\begin{tabular}{ccccccccccc}
\multicolumn{11}{c}{\footnotesize {\bf Table 1:} Maser feature parameters} \\
& & & & & & &  & & & \\
\hline\hline
\multicolumn{1}{c}{} & \multicolumn{1}{c}{Feature} & \multicolumn{1}{c}{$V_{\rm LSR}$} &
\multicolumn{1}{c}{$F_{\rm int}$} &  &
\multicolumn{1}{c}{$\Delta \alpha$} & \multicolumn{1}{c}{$\Delta \delta$} &
& \multicolumn{1}{c}{$V_{\rm x}$} & \multicolumn{1}{c}{$V_{\rm y}$} &
\multicolumn{1}{c}{$V_{\rm mod}$} \\
\multicolumn{1}{c}{} & & \multicolumn{1}{c}{(km s$^{-1}$)} &
\multicolumn{1}{c}{(Jy)} &   &
   \multicolumn{1}{c}{(mas)} & \multicolumn{1}{c}{(mas)} &
   & \multicolumn{1}{c}{(km s$^{-1}$)} & \multicolumn{1}{c}{(km s$^{-1}$)} &
   \multicolumn{1}{c}{((km s$^{-1}$)} \\
\hline
 & 1        &  --7.2 & 16.9 &  &   189.7 (0.3) &  1020.7 (0.2) & & {\itshape 23.7 (12.6)} $^{\dag}$  &  {\itshape --33.8 (8.2)} & {\itshape 41.2 (9.9)} \\
 & 2        &  --4.8 & 5.9 &      &   157.7 (0.2) &  1010.43 (0.03) & & 4.2 (1.8) &  --15.2 (0.4) &  15.8 (0.6) \\
 & 3        & --4.8 & 0.5 &  &   157.3 (0.2) &  1012.1 (0.2) & &  &    &   \\
Group I & 4        &  --3.6 & 0.8 &  &   159.0 (0.2) &  1017.26 (0.03) & &  &    &   \\
 & 5        &  --2.0 & 4.1 & &   --32.5 (0.2) &  1290.15 (0.09) & &  &   &   \\
 & 6        &  --1.2 & 1.0  & &   174.3 (0.3) &  1007.5 (0.2) & & 4.4 (2.7) &  --19.1 (1.3) &  19.6 (1.4) \\
 & 7        &  --0.6 & 0.3 & &   28.7 (0.2) &  1251.03 (0.07) & &  &    & \\
 & 8        &  0.7 & 0.3 & &   55.5 (0.2) &  1245.85 (0.08) & &  &    &   \\
& & & & & & &  & & & \\
 & 9        &  --5.4 & 0.3 & &   --50.3 (0.3) &  --18.99 (0.02) & &  &    &   \\
 Group II & 10        &  --5.2 & 4.8 & &     0.00 &     0.00 & & 0.00 &  0.00 &  0.00 \\
 & 11        &  --3.9 & 3.0 &  &  --655.2 (0.3) &  --566.7 (0.1) & & {\itshape --25.3 (15.0)} &  {\itshape --9.8 (4.8)} &  {\itshape 27.1 (14.1)} \\
  & 12       &  --3.7 & 1.2 &  &   --179.3 (0.2) &  --75.6 (0.1) & & --1.0 (1.9) & 0.6 (1.3)    & 1.2 (1.8)  \\
 \hline
\end{tabular}
%
%
\begin{flushleft}
${\dag}$ The italics indicate tentative values of proper motion components for  features observed at only two epochs
\end{flushleft}
\end{table*}

AFGL 5142 was observed in the \(6_{16}-5_{23}\) H$_2$O maser line (rest frequency 
22235.080~MHz) using the European VLBI Network (EVN) at four epochs (October
1996, and June, September, November 1997), each epoch consisting of 13 scans
(of 6.5 minutes) distributed over an hour angle of 11 hours. The antennae 
involved in the observations were Medicina, Cambridge, Onsala, Effelsberg, 
Metsahovi, Noto, Jodrell and Shanghai\footnote{The baselines to Shanghai have been almost completely flagged during the process of self-calibration.}, with a subset of  5--7 of these (including always Effelsberg) observing at each epoch. 
Before and after each scan on AFGL 5142, several continuum sources were observed 
for calibration purposes (0528+134, 0642+449, 1803+784, 2251+158, 3C454.3, OJ287). 
Both circular polarizations were recorded with a 1~MHz bandwidth centered at the LSR
velocity of --4.8~km~s$^{-1}$. The data were processed with the MKIII correlator
at the Max-Planck-Institut f{\"u}r Radioastronomie (Bonn, Germany), obtaining 112 
spectral channels with a separation of 0.12~km~s$^{-1}$.

Data reduction was performed using the NRAO AIPS package, following the 
standard procedure for VLBI line data. 
Total power spectra of continuum calibrators were used to derive the bandpass 
response of each antenna. 
The amplitude calibration was performed using the information on the system 
temperature and the gain curve of each antenna. We did not use the 
``template spectrum'' method (which consists of comparing total power spectra of 
different scans) owing to insufficient signal-to-noise ratio (SNR) of
the total power spectra of the maser source. 

For each observing epoch, a single scan of a strong calibrator was used to derive the 
instrumental (time-independent) single-band delay  and the phase offset between the 
two polarizations. 
After removing the instrumental errors, all calibrator scans were fringe-fitted to 
determine the residual (time-dependent) delay and the fringe rate. The corrections 
derived from calibrators were applied to the strong maser feature used
as a phase reference.

In the data analysis of each observing epoch we used as phase-reference the
 same maser feature at $V_{\rm LSR}$ = --4.8~km~s$^{-1}$, which exhibits a simple spatial 
structure consisting of a single, almost unresolved spot. 
The visibilities of the phase-reference channel were fringe-fitted to find the 
residual fringe rate produced both by differences in atmospheric fluctuations between 
the calibrators and the source, and by errors in the model used at the correlator. 
After correcting for the residual fringe rate, the visibilities of the reference channel were 
self-calibrated to remove any possible effects of spatial structure. Finally, the corrections derived 
from the reference channel were applied to data of all spectral channels. 

Before producing synthesis aperture maps, we searched for maser emission in a 
wide field ($12''\times12''$) centered on the reference feature using the 
fringe-rate mapping method \citep{Wal81}. The detected emission centers,
all within a few arcseconds from the phase center,
were conveniently mapped by producing
(naturally weighted) maps extended over a sky area of $(\Delta \alpha \  
cos\delta \times \Delta \delta) \ 4''\times 4''$ and covering
the velocity range from \ $-$10.5 to 0.7~km~s$^{-1}$. 
The CLEAN beam was an 
elliptical gaussian with a typical FWHM size of $ 2.1 
\times 1.1 $ mas. In each observing epoch, the RMS noise level on the channel maps, $\sigma$, is close to the theoretical thermal value, 0.03~Jy~beam$^{-1}$, for channels where no
signal is detected, and increases to \ 0.3~Jy~beam$^{-1}$ for channels with the
strongest components.

Every channel map has been searched for emission above a conservative 
detection threshold 
(in the range 5--10~$\sigma$), and the detected maser spots have been fitted with two-dimensional elliptical Gaussians, determining  position, flux density, and FWHM size of the emission.  
Hereafter, we use the term of ``feature'' to indicate a collection of spectrally and spatially contiguous maser spots.
A maser feature is considered real if it is detected in at least three contiguous channels, with a position shift of the intensity peak from channel to channel smaller than the FWHM size. 
Fig.~\ref{spec} compares the Effelsberg total-power spectrum observed in Oct 1996 with the integrated flux densities of the VLBI channel maps for the same epoch. One notes that, excepted for the velocity range
from \ $-$7 to $-$8 ~km~s$^{-1}$, the difference between the total-power flux and the flux recovered in the imaged VLBI field of view is within the amplitude calibration errors ($\pm$ ~50\%).

The relative positional uncertainty of the single maser spot is estimated using the expression
\begin{equation}   
\Delta \theta = \frac{\sigma }{2 \, I } \; FWHM \,,
\end{equation}
where $FWHM$ is the un-deconvolved spot size, $I$ is the peak intensity and
$\sigma$ is the no-signal rms of the map 
\citep{Rei88}.
For most of the spots the positional uncertainty is of the order of 
$\sim$50 -- 100~$\mu$as.
%
\section{Observational Results}

%
\begin{figure*}
\centering
\includegraphics[width = 18cm]{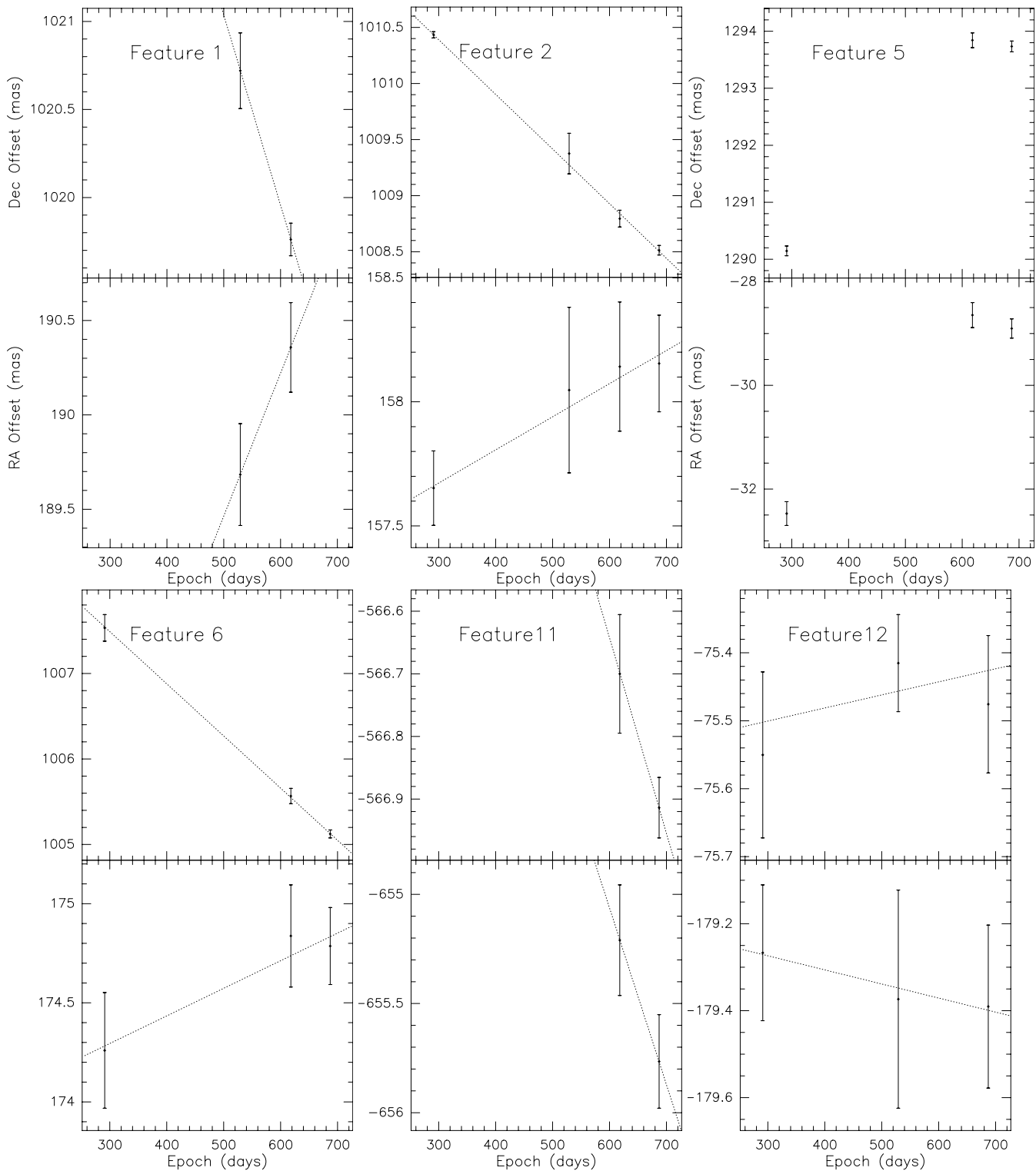}
\caption{Measured relative proper motions of H$_2$O maser features in AFGL 5142.
For each of the time persistent features, the top and the bottom panels report the 
time variation of respectively the declination and the right ascension offsets (relative to the
feature "10").
In each panel, the dotted line shows the proper motion calculated by the 
(error-weighted) linear least-squares fit of the positional offsets with time. Among the features observed at three or more epochs,  the proper motions are derived
 only for those moving in a straight line at constant velocity (within the positional errors). Adopting such
 a criterion, no proper motion is derived for  feature "5". Tentative 
 values of the proper motions are calculated also for the two features 
 (label number "1" and "11") observed at only two epochs.}
\label{comp_prmot}
\end{figure*}

Counting all four epochs, 26 H$_2$O maser features were detected. 
Several of these show a good agreement in relative positions (within few mas) and line of 
sight velocities (within 1 km s$^{-1}$) for two or more epochs, and  
therefore we assume that in these cases we identified features which persist over time. 
A final set of 12 distinct features, 7 out of these 
observed for more than one epoch, has been identified.

Table~1 gives the parameters of the features, determined by fitting an 
elliptical Gaussian to the intensity distribution of the maser spots
contributing to the features' emission at different velocity channels.  
Col.~1 gives the feature label number. Cols.~2  and~3 report 
respectively the line-of-sight
velocity, $V_{\rm LSR}$, and  the integrated flux density, $F_{\rm int}$,  of the highest-intensity channel, both averaged over the 
observational epochs for the time-persistent features. 
For each feature,
Cols. 4 and 5 report the positional offset (of the first epoch of
detection) calculated with respect to the 
feature with label number ``10'', present at all four epochs. 
The positional offsets of a given feature are estimated from the (error-weighted) 
mean  positions of the contributing maser spots. 
The bracketed numbers are 
the relative positional uncertainties, evaluated by taking the weighted standard 
deviation of the spot positions.
With the \ 1--2~mas angular resolution of the EVN at 22~GHz, the emission of
most of the spots is found to be spatially unresolved. Using a distance value
of 1.8~kpc for AFGL 5142, an upper limit to the spot size of \ $\sim$2~AU \ is 
derived.

The proper motions have been calculated  performing a (error-weighted) linear 
least-squares fit of the positional offsets with time. Fig.~\ref{comp_prmot} shows the 
time variation of the right ascension and declination offsets (relative to the
feature ``10'') for the persistent features. Among the features observed
at three or more epochs,  the proper motions are derived
only for those moving in a straight line at constant velocity (identified with 
the label numbers ``2'', ``6'' and ``12''). Tentative values of the proper motions
are calculated also for the two features (with label number ``1'' and ``11'') 
observed
at only two epochs. Cols. 6, 7 and 8 of Table~1 report respectively the 
projected components along the R.A. and DEC axis, and the absolute value of the derived proper motions.
The numbers in italics refer to features observed at only two epochs.
The bracketed numbers are the formal errors of the linear least-squares fit.

\begin{figure*}
\centering
\includegraphics[width=13.5cm]{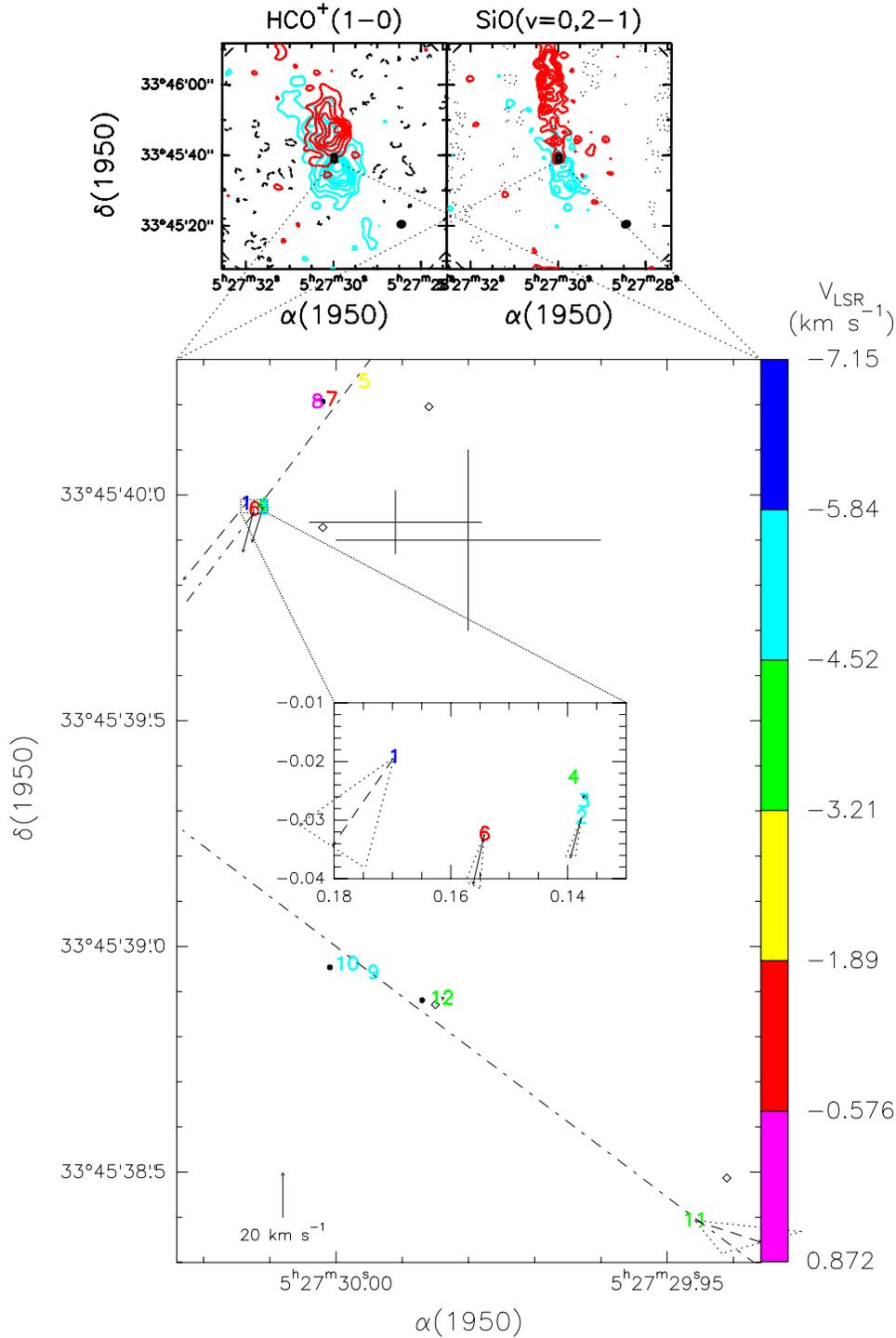}

\caption{The upper panels show the contour maps of the high-velocity molecular 
line emission of HCO$^{+}$ (1 $\to$ 0) ({\em left}) and SiO 
($v = 0$, 2 $\to$ 1) ({\em right}), observed with OVRO in 1998 by Hunter et al. (1999; their Fig. 10). 
 The area at the center of the OVRO field of view, where  most
 of the 22.2 GHz 
water emission detected by our VLBI observations is concentrated, 
is expanded in the lower panel. This shows the spatial distribution
of the VLBI features  superimposed on top of the distribution of the VLA 
emission centers found 
by \citet{Hun95,Hun99} at the two different epochs 
(1992, open diamonds;  1998, filled dots).
Each VLBI feature is identified with the label number given in 
Col.~1 of Table 1. Different colours are used to distinguish the line of 
sight velocities of the features, according to the colour-velocity conversion
code shown on the right-hand side of the panel. The arrows indicate the measured
proper motions, with dashed lines used in case of more uncertain values. 
The dotted triangles drawn around the proper motion vectors represent the
 amplitude and orientation uncertainty.  
The region in the northeast corner enclosed by the dotted lines  is shown in greater detail in the box at the center of the lower panel, whose coordinates are arcsec offsets relative to the position: RA = $05^h 27^m 30^s$; DEC = +$33^\circ 45' 40''$. 
For the feature labeled ``12'', with 
a  small measured value of the proper motion, cannot be assigned
 a definitive direction of motion.
The amplitude scale for the proper motions is given at the bottom left corner 
of the panel.
The small and the big crosses give respectively the position uncertainties of the VLA 8.4 GHz and the OVRO 88 GHz continuum emission peaks \citep{Hun99}.
For each of the two Groups of EVN maser features, the black dot-dashed line indicates the axis whose average distance from the feature positions is minimum.}
\label{prmot}
\end{figure*}

Fig.~\ref{prmot} compares our VLBI results with previous 
interferometric observations.
Top panels show the high-velocity molecular outflows seen in HCO$^{+}$ 
(1 $\to$ 0)  and SiO ($v = 0$, 2 $\to$ 1) with OVRO 
\citep{Hun99}. The area comprising almost the totality of the 22.2 GHz 
water maser 
emission detected by our VLBI observations, is indicated by a small 
filled rectangle at the center of the field of view. This area is expanded in 
the lower panel of Fig.~\ref{prmot}, which shows the spatial distribution
of the VLBI maser features superimposed on top of the distribution of the 
VLA emission centers found by \citet{Hun95,Hun99} 
at two different epochs. The measured proper motions are indicated
by the arrows.
 
The absolute position of the VLBI map has been determined by aligning the 
VLBI and VLA maser emission centers.  The VLBI map has been
shifted on top of the VLA map to find a shift that minimizes the 
root mean square difference between the positions of the VLBI and VLA spots. 
We find that each of the detected VLBI features has a good positional 
(within the VLA beam
of $\sim$100~mas) and line-of-sight velocity correspondence with one of the
observed VLA spots. Conversely, we have detected VLBI features corresponding to
 each VLA emission center with peak flux density $\geq$ 0.4~Jy. The good
overlap found between the VLBI and the VLA emission makes us  confident
that the absolute position derived for the VLBI map is accurate within
the VLA positional uncertainty.


On the basis of their spatial 
distribution, the VLBI maser features can be divided into two groups. Group I, 
comprising the first eight features,
is found in the northeast corner of the area plotted in the lower panel of 
Fig.~\ref{prmot}, in the same region (within $\sim$500~mas) where the 
8.4~GHz and 88~GHz continuum emissions are detected.
Group II includes the last four maser features 
more detached  ($\geq$ 1$''$) from the continuum emissions, extending towards 
the south and the southwest corner of the plotted area. 

Fig.~3 and Table~1 of \citet{Hun99} show that at epoch 1998 two VLA emission centers (their components ''6'' and ''7'') are detected at positions separated about 3--4$^{''}$ from the 8.4~GHz continuum peak. We produced VLBI maps also at these two locations, toward both directions detecting a single, weak 
($\leq$Jy beam$^{-1}$) spot in the last two epochs (September and November 97). 
No features persisted over time and, consequently, no proper motions are measured. 
The components
''6'' and ''7'' were among the strongest ones in the epoch 1998 VLA observations of
\citet{Hun99} but were not detected in the two prior VLA epochs (1991, \citealt{Tor92}; 1992, \citealt{Hun95}). Our VLBI observations, extending over the years 1996 and 1997, fall in between the 1992
\citep{Hun95} and 1998 \citep{Hun99} VLA runs, and witness the first appearance of water maser emission in these two regions. Being so detached from the 8.4~GHz and 88~GHz sources, the maser emission in these two locations very likely traces
a site of star formation other than the one responsible for the line and continuum emission shown in Fig.~\ref{prmot}. A single maser spot does not
allow us to derive information on the gas kinematics and we will not consider 
further these two regions in the following discussion.

%
\section{Discussion}


Fig.~\ref{prmot} shows that the spatial distribution of all the VLBI and VLA maser features is extended along a north$-$south direction. The outflow detected in the CO, HCO$^+$ and SiO emission has a similar orientation \citep{Hun95,Hun99}. 
In addition, although the line-of-sight velocities of the VLBI maser features do not vary smoothly, the mean velocity of the Group I cluster (toward the north) is more positive (redshifted) than that of the Group II (toward the south).
 Over an area of sky sligthly larger than that plotted in Fig.~\ref{prmot}, a similar variation of the line-of-sight velocities (more redshifted (blueshifted) toward the North (South)) is also noteable in the VLA 22~GHz observations of
Hunter et al. (1999; their Table~1), where the most northward VLA maser components (labeled ''9'', ''10'' and ''11'' in \citealt{Hun99}) occur at velocities ($\geq$ --1~km~s$^{-1}$)  higher than the most southwarth components (labeled ''2'' and ''8'' in \citealt{Hun99} ) ($\leq$ --3~km~s$^{-1}$). The observed variation of water maser velocity 
is qualitatively in agreement with the velocity distribution seen at a much larger angular scale in the HCO$^+$ and SiO maps (Fig.~\ref{prmot}).
A simple interpretation might be that all the detected (VLA and VLBI) maser features are tracing the flow motion in the innermost portion of the molecular outflow. To make
the discussion a bit more quantitative, considering that the line-of-sight velocity 
dispersion of the large-scale (diameter $\sim$ 50$''$) molecular outflow is 
\ $\sim$100~km~s$^{-1}$, and assuming a Hubble flow (velocity increasing linearly along the outflow axis), one would derive a rate of line-of-sight velocity 
dispersion caused by the outflow of \ $\sim$2~~km~s$^{-1}$~arcsec$^{-1}$. If 
that
might explain the velocity dispersion of the Group II features (1.7~km~s$^{-1}$
over a distance of \ 1$''$; see Table~1), the Group I features show  
a much higher velocity dispersion (8~km~s$^{-1}$) across a smaller distance
(0\pas35).

The maser features of Group I have a sky-projected distance \ $\leq$ 500--1000~AU \ from the 8.4~GHz and 88~GHz
continuum sources, and should emerge near to the expected location
of the massive YSO. At such close distance from the YSO, it might be possible
that the flow motion has not yet reached a stable configuration and 
turbulence might play a role in increasing the velocity dispersion of the gas.
Alternatively, the maser features of Group I might move under the influence of 
the gravitational field of the massive YSO.
Within a region of radius of \ $\sim$1000~AU around the 
forming high-mass star, the current theory of star formation predicts that an
accretion disk should be found.
Looking at Fig.~\ref{prmot}, one notes that the maser features of Group I  
have an elongated spatial distribution (the dot-dashed line indicates the elongation axis) and that the measured proper motions have orientation close to 
that of the elongation axis. This geometrical condition is what one would 
in principle expect if these maser features traced a rotating disk seen edge-on.
However, one should note that the derived proper motions are relative to the
feature ''10'' (not belonging to the Group I maser cluster), and, in order to obtain the absolute transversal velocities of the gas, one has to correct for 
the (unkwown) absolute motion of this feature.


Even if only a small number of maser features is detected toward the Group I cluster, neverthless our accurate knowledge of their positions and line-of-sight
velocities offers the chance to fit their motion using a 3-D Keplerian disk 
model. 
The model's free parameters are: the 
sky-projected coordinates of the YSO (at the disk center); the position angle
and the inclination angle with the line-of-sight of the disk axis; the YSO
mass. For a given set of input parameters, one can compute the position
and the velocity vector of each feature and compare the model velocities
with the observed velocities. The best fit to the data was obtained
minimizing the \ $\chi^{2}$ \ given by the squared sum of the error-weighted
differences of the model and observed line-of-sight velocities.
Incidentally we note that the same fit solution is found when the \ $\chi^{2}$ \ is 
calculated including also the two measured proper motions (for the features 
labelled ''2'' and ''6'' in Table~1). 

Looking for the disk axis orientation over the full \ 4$\pi$ \ solid angle, the
best fit solution is found with the disk seen almost edge-on (inclined 12$^{o}$
from the line-of-sight) and oriented on the sky parallel
with the elongation axis of the Group I features (at P.A. = 153$^{o}$).
The fitted value of the YSO mass, $M_{YSO} = 38~M_{\odot} \pm 20~M_{\odot}$,
although determined with high uncertainty, strongly indicates
 that the central object is a massive YSO ($M > 10~M_{\odot}$).
This result is in agreement with that of previous observations \citep{Hun95, Hun99}, indicating an exciting object of spectral type B2 or earlier, for which the theoretically expected value of the mass is $\geq 10~M_{\odot}$ \citep{Vac96, Pal02}.  
Adopting a distance of 1.8~kpc to AFGL 5142, the 
range of disk radii traced in our model by the maser emission extends 
from $\sim$30~AU
to $\sim$800~AU, which is consistent with the size of several hundreds of AU
expected for an accreting disk around a massive YSO. 
From the barely resolved 88~GHz emission, interpreted in terms of optically 
thin thermal emission from a dusty core, \citet{Hun99} derive a core size of \ 
$\sim$5000~AU and a core mass of \ $\sim$145~M$_{\odot}$. These values are 
compatible with the results of our model, which tells us that within a radius
of \ $\sim$1000~AU \ the mass in Keplerian motion is \ $\lesssim$ 60~M$_{\odot}$.

 Recently \citet{Zha02} have observed the high-mass star-forming region AFGL 5142 in several NH$_{3}$ inversion transitions using the VLA array. In correspondence of the 88~GHz source of \citet{Hun99}, they find a compact (1$''$ in diameter), hot (70~K), NH$_{3}$ structure, with a broad line emission, interpreted as an
 unresolved rotating disk. Looking at Fig.~2 of \citet{Zha02}, one sees that the velocity range (from --8~km~s$^{-1}$ to 2~km~s$^{-1}$) over which the NH$_{3}$ "disk" emission is detected matches well with the velocity range of the Group I maser emission. The value of the "disk" mass 
 estimated from the NH$_{3}$ measurements of \citet{Zha02} is \ 4~M$_{\odot}$, assuming an NH$_{3}$ abundance (relative to 
 H$_{2}$) of \ 10$^{-6}$. This value of the "disk" mass, being much lower than the fitted YSO mass (38~M$_{\odot}$), is compatible with the
 Keplerian disk model proposed to  explain the kinematics of the 22 GHz maser features that we observed.

Owing to their large distance (2000--3000~AU) from the YSO believed to be responsible for the continuum emissions and the
acceleration of the large scale molecular ouflow, the Group II of maser features might be also associated with a distinct (as yet undetected) YSO.
However, the fact that the positions and the radial velocities of
these features are in agreement with the blue-shifted lobe of the
molecular outflow, makes us favour the intepretation that their emission is excited by the 
interaction of the gas outflowing from the YSO with the ambient gas of the progenitor molecular core. Looking at Fig.~\ref{prmot} one notes that the spatial
distribution of the maser features of Group II is elongated (the dot-dashed
line indicates the elongation axis) along a direction (at P.A. = 44$^{\circ}$)
that differs from that of the large scale HCO$^{+}$ and SiO outflows, 
oriented approximately north-south. Towards the infrared
sources IRS1 and IRS3 in the NGC2071 star-forming region, \citet{Set02} note
similar differences of orientation between the outflow structures
on large ($>$ 1000~AU) and small (10--100~AU) scales. This effect might be
explained thinking of either the density gradients in the
ambient medium, causing large-angle bends of the protostellar jets,
or multiple, small scale outflows, whose merging creates the large
scale flow. 
%
\section{Conclusions}

Using the EVN we have observed the 22.2 GHz H$_2$O 
masers towards the massive star forming region AFGL 5142 for four epochs (from October 1996 to November 1997). Previous high-angular resolution observations of
several thermal tracers in the cm and mm-wavelength band indicate that 
the exciting source of the water maser emission is a high-mass YSO, of
spectral type B2 or earlier.

We identified a final set of 12 distinct water maser features, 7 out of 
these detected in more than one epoch.
For a few features, persistent over three
or four epochs, accurate values of the proper motions are derived. The
observed proper motions have amplitudes of 15--20~km~s$^{-1}$, significantly
larger than the range of variation of the line-of-sight velocities ($\pm$4~km~s$^{-1}$ around the systemic velocity).
The positions and the velocities of the VLBI features agree well with those
of the emission centers previously identified by means of VLA observations.

On the basis of their spatial distribution, we have divided the maser features into two groups. Group I, 
comprising eight VLBI features,
is found in the same region (within $\sim$500~mas) where the massive
YSO, believed to be the exciting source of two 
compact, continuum sources (at 8.4 GHz and 88 GHz), should be located.
A model fit to the positions and velocities of these  features indicates that 
they might arise on the surface of a  Keplerian disk seen nearly edge-on.
The fitted value of the YSO mass, $M_{YSO} = 38~M_{\odot} \pm 20~M_{\odot}$,
although determined with high uncertainty, is 
in agreement with the results of  previous observations.
Maser features of Group II, found at larger 
distances ($\geq$ 1$''$) southward from the YSO, have positions and
line-of-sight velocities in agreement with the blue-shifted lobe
of a large scale molecular outflow (detected in the HCO$^{+}$ and SiO
emission), and
might be excited by the interaction of the
gas outflowing from the YSO with the ambient gas of the progenitor molecular
core.

This work demonstrates that multi-epoch EVN observations are able to measure
the proper motions of the strongest and longer-living  22.2~GHz maser features.
Our EVN observations towards the source AFGL 5142 suffered two major drawbacks:
\ 1) the time separation between consecutive epochs ($\geq$ 3~months) is too long 
compared to the average life time of the maser features; \ 2) modest sensitivity
 (with an average detection threshold of $\sim$0.35~Jy~beam$^{-1}$), owing to the fact that
 only 5--7 antennae (out of the 11 available to observe at 22.2~GHz) took part in each run.
Next we plan to use the VLBA to better constrain
the kinematical scenario suggested by the EVN observations, using a shorter
time separation ($\sim$1~month) between two consecutive epochs and higher
sensitivity.

\begin{acknowledgements}
      The European VLBI Network is a joint
facility of European, Chinese, South African and other radio astronomy
institutes funded by their national research councils.

We are very gratful to L. Testi for providing the plots of the molecular outflows.
\end{acknowledgements}

\bibliographystyle{aa}
\bibliography{biblio.bib}

\end{document}